\documentclass[3p,times,procedia]{elsarticle}
\usepackage{nupha_ecrc}

\volume{00}

\firstpage{1}

\journalname{Nuclear Physics A}

\runauth{}

\jid{nupha}

\jnltitlelogo{Nuclear Physics A}

\usepackage{amssymb}
\usepackage[figuresright]{rotating}

\begin{document}

\begin{frontmatter}



%% Instructions from Editor: Please use the following \dochead only in the preprint version (e-print arXiv etc.); 
%% use empty \dochead{} when submitting to Nuclear Physics A!
\dochead{XXVIIth International Conference on Ultrarelativistic Nucleus-Nucleus Collisions\\ (Quark Matter 2018)}
%% e.g. \dochead{17th International Conference on Dynamical Processes in Excited States of Solids}

\title{Latest predictions from the EbyE NLO EKRT model}


\author[a,b]{H.~Niemi}
\author[a,b]{K.~J.~Eskola}
\author[b,c]{R.~Paatelainen}
\author[b,c]{K.~Tuominen}

\address[a]{University of Jyv\"askyl\"a, Department of Physics, P.O.~Box 35, FI-40014 University of Jyv\"askyl\"a, Finland}
\address[b]{Helsinki Institute of Physics, P.O.Box 64, FI-00014 University of Helsinki, Finland}
\address[c]{Department of Physics, University of Helsinki, P.O.~Box 64, FI-00014 University of Helsinki, Finland}

\begin{abstract}
We present the latest results from the NLO pQCD + saturation + viscous hydrodynamics (EbyE NLO EKRT) model. The parameters in the EKRT saturation model are fixed by the charged hadron multiplicity in the 0-5 \% 2.76 TeV Pb+Pb collisions. The $\sqrt{s}$, $A$ and centrality dependence of the initial particle production follows then from the QCD dynamics of the model. This allows us to predict the $\sqrt{s}$ and $A$ dependence of the particle production. We show that our results are in an excellent agreement with the low-$p_T$ data from 2.76 TeV and 5.02 TeV Pb+Pb collisions at the LHC as well as with the data from the 200 GeV Au+Au collisions at RHIC. In particular, we study the centrality dependences of hadronic multiplicities, flow coefficients, and various flow correlations. Furthermore, the nuclear mass number dependence of the initial particle production and hydrodynamic evolution can be tested in the 5.44 TeV Xe+Xe collisions at the LHC. To this end, we show our predictions for charged particle multiplicities, and in particular, show how the deformations of the Xe nuclei reflect into the flow coefficients.    
\end{abstract}

\begin{keyword}
heavy-ion collisions \sep perturbative QCD calculations \sep saturation \sep dissipative fluid dynamic
\end{keyword}

\end{frontmatter}

\section{Introduction}
\label{sec:intro}

In the EKRT framework the initial conditions for hydrodynamical evolution in heavy-ion collisions are calculated by using NLO perturbative QCD and collinear factorization together with a saturation conjecture that controls the transverse energy production through a semi-hard scale $p_{\rm sat}$ \cite{Eskola:1999fc, Paatelainen:2012at}. The saturation momentum is a function of $\sqrt{s}$ and the nuclear mass number, and depends on the transverse coordinate through the product of nuclar thickness functions $T_A$ \cite{Paatelainen:2013eea,Eskola:2001rx},
% \begin{equation}
 $p_{\rm sat} = p_{\rm sat}(T_A T_A, \sqrt{s}, A, K_{\rm sat})$, 
% \end{equation}
where $K_{\rm sat}$ is fixed by the charged particle multiplicity in $0-5$ \% 2.76 TeV Pb+Pb collisions. Once $K_{\rm sat}$ is fixed, the initial conditions can be computed for any $\sqrt{s}$ and $A$, as long as they are sufficiently large, so that the $p_{\rm sat}$ remains perturbative.

The local formation time can be estimated as $\tau_s(\mathbf{r}) = 1/p_{\rm sat}(\mathbf{r})$ and the local energy density at the formation time is then given by
% \begin{equation}
$e(\mathbf{r},\tau_{\mathrm{s}}(\mathbf{r})) = \frac{K_{\rm sat}}{\pi}[p_{\rm sat}(\mathbf{r})]^4$ \cite{Eskola:2001bf}.
% \end{equation}
This profile is then evolved to a common time $\tau_0=1/p_{\rm sat}^{\rm min}=0.2$~fm by using 0+1D Bjorken hydrodynamics. The evolved energy density profile can then be used as an initial condition for the full fluid dynamical evolution. 

The subsequent evolution of the system is described by a boost-invariant Israel-Stewart type of dissipative fluid dynamics with the coefficients of the non-linear terms from Refs.~\cite{Denicol:2012cn, Molnar:2013lta}. 
% The equation of state is the $s95p$-PCE-v1 parametrization~\cite{Huovinen:2009yb}, kinetic freeze-out is $T_{\rm dec}=100$ MeV, and the chemical freeze-out is at $T_{\rm chem} = 175$ MeV.
The kinetic freeze-out is $T_{\rm dec}=100$ MeV. The chemical freeze-out at $T_{\rm chem} = 175$ MeV is encoded into the equation state, for which we use the $s95p$-PCE-v1 parametrization~\cite{Huovinen:2009yb, Huovinen:2007xh}. 
We neglect the bulk viscosity and set the initial values of shear-stress tensor and transverse velocity to zero. The remaining free input is the parametrization of the temperature dependence of the shear viscosity to entropy density ratio, $\eta/s(T)$. 

As discussed in detail in Ref.~\cite{Niemi:2015qia}, the nuclear shapes and the event-by-event fluctuations enter the calculation through $T_A$. The event-by-event $T_A$ is computed by sampling the nucleon positions from the Woods-Saxon parametrization of the nucleon density. For each nucleon we then set a gaussian transverse density profile, and the nuclear $T_A$ is the sum of these nucleon thickness functions. We take the Pb and Au nuclei to be spherically symmetric, but as in Ref.~\cite{Giacalone:2017dud}, for the Xe nuclei we take into account the nuclear shape deformation, described by the parameters $\beta_s = 0.163$ and $\beta_4 = -0.003$ from Ref.~\cite{Moller:2015fba}.

\section{Results}
\label{sec:results}

%%%%%%%%%%%%%%%%%%%%% FIGURE %%%%%%%%%%%%%%%%%%%%%
\begin{figure}
\begin{center}
\includegraphics[width=0.32\textwidth]{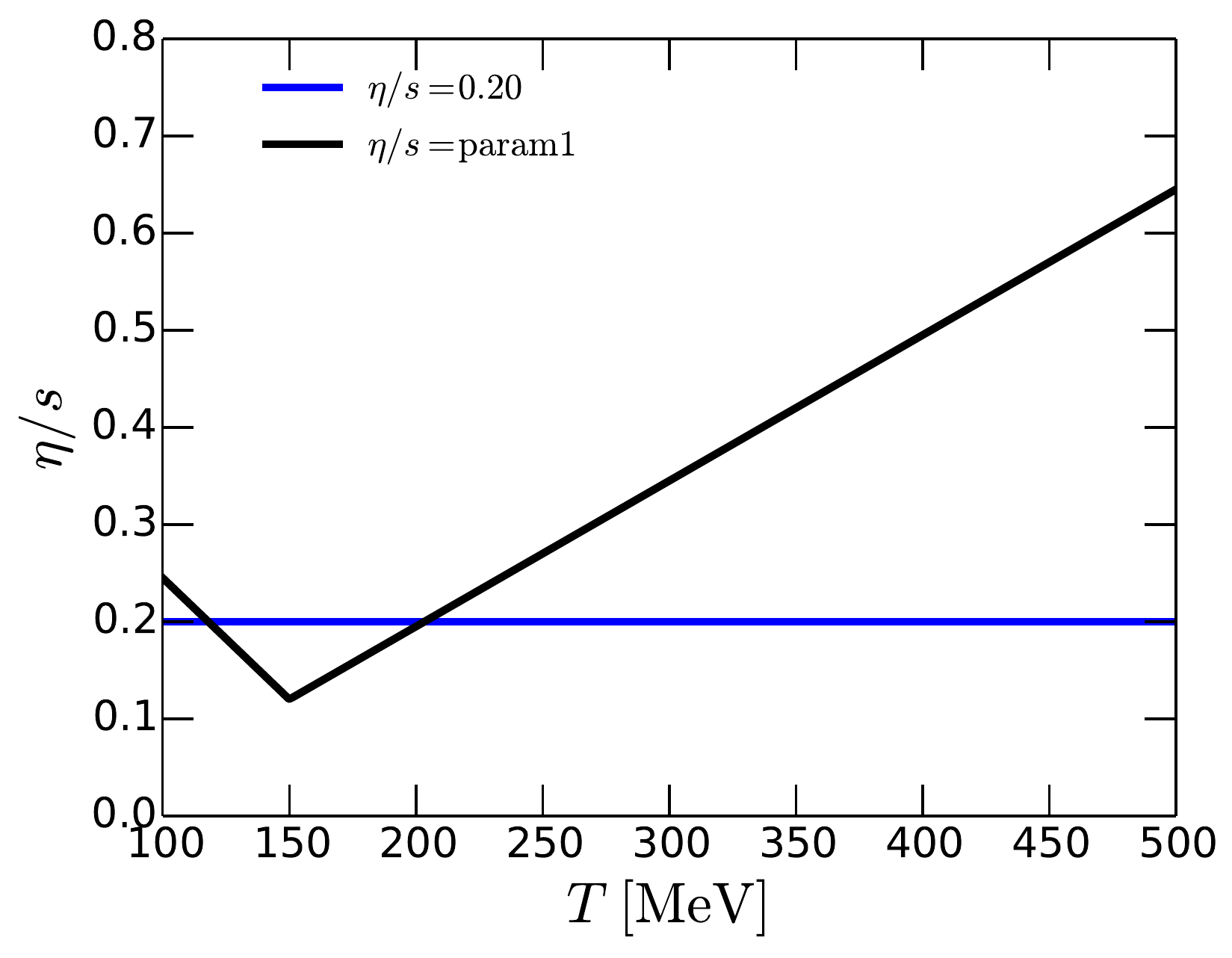}
\includegraphics[width=0.32\textwidth]{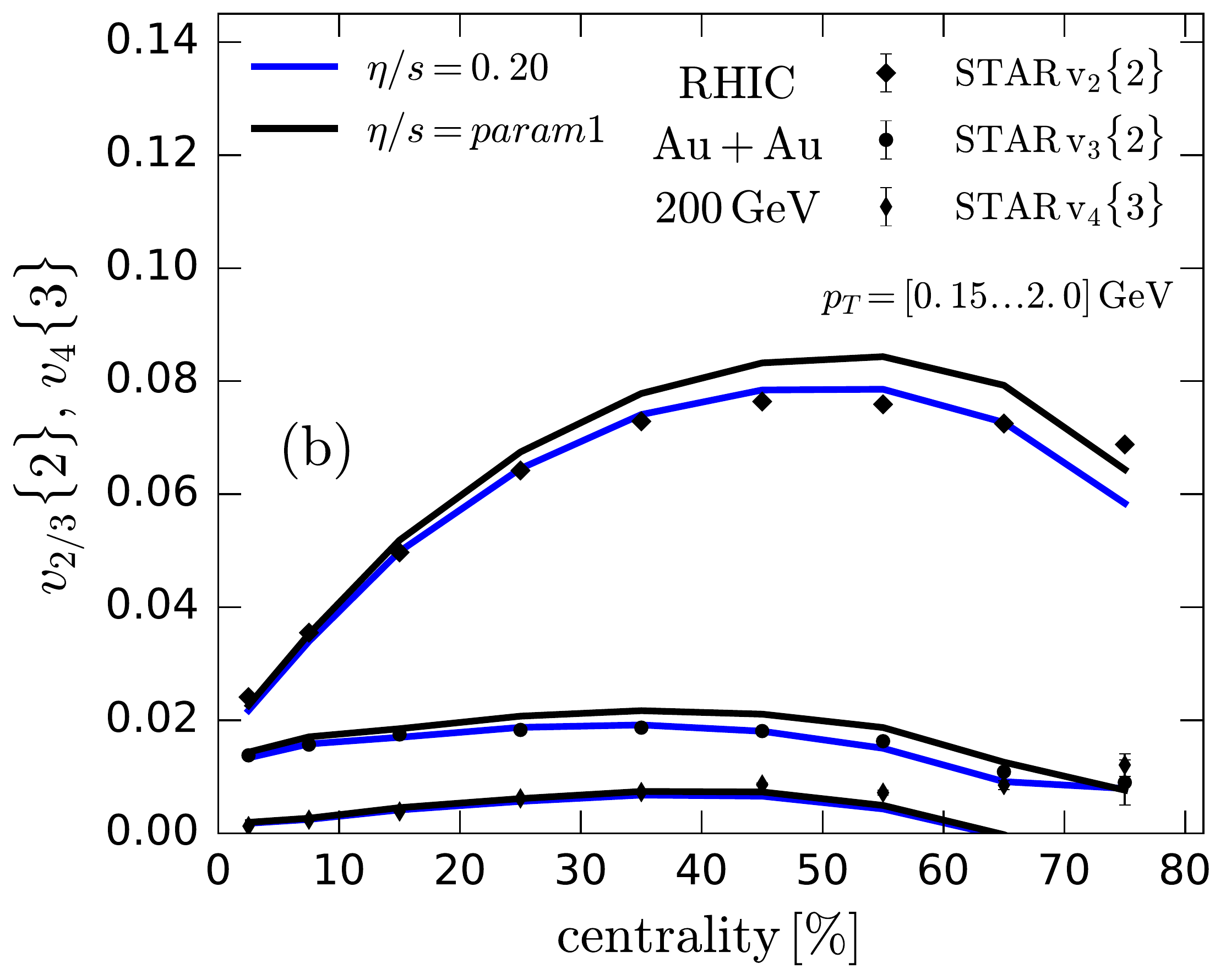}
\includegraphics[width=0.32\textwidth]{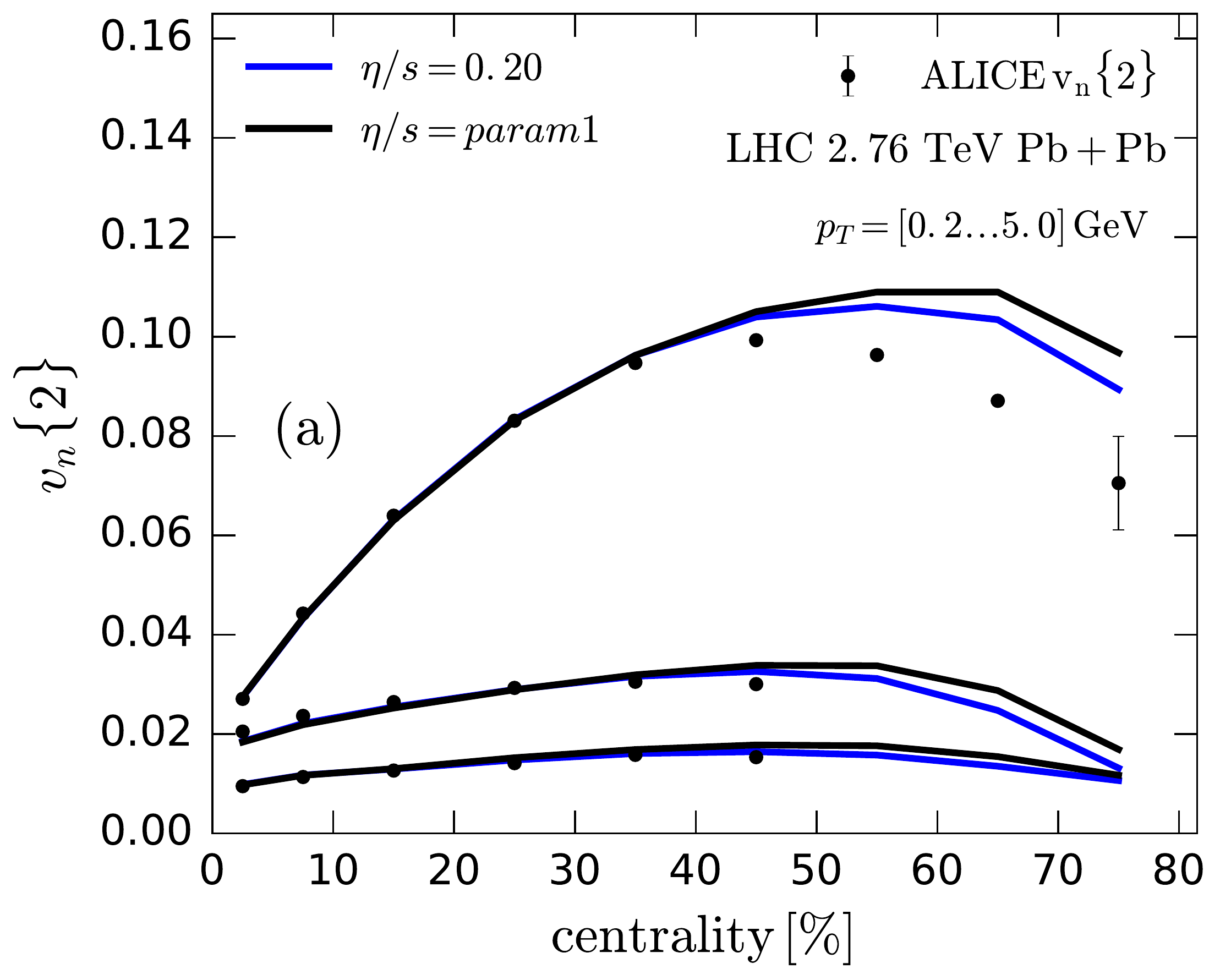}
\end{center}
\vspace{-0.5cm}
\caption{\small The tested temperature dependencies of $\eta/s$ (Left). The centrality dependence of $v_n$ in 200 GeV Au+Au collisions (Middle), and 2.76 TeV Pb+Pb collisions (Right). The experimental data are from ALICE \cite{ALICE:2011ab} and STAR \cite{Adams:2004bi, Adamczyk:2013waa, Adams:2003zg}}
\vspace{-0.3cm}
\label{fig:etapers}
\end{figure}
%%%%%%%%%%%%%%%%%%%%% FIGURE %%%%%%%%%%%%%%%%%%%%%
The parametrizations of $\eta/s(T)$ are constrained by the flow coefficients $v_n$ and various flow correlators in 200 GeV Au+Au and 2.76 TeV Pb+Pb collisions \cite{Niemi:2015qia}. The $\eta/s(T)$ parametrizations that give the overall best agreement with these data are shown in Fig.~\ref{fig:etapers} (Left), and the corresponding $v_n$ are shown in Figs.~\ref{fig:etapers} (Middle) and \ref{fig:etapers} (Right), respectively. 

\begin{figure}
\begin{center}
\includegraphics[width=0.95\textwidth]{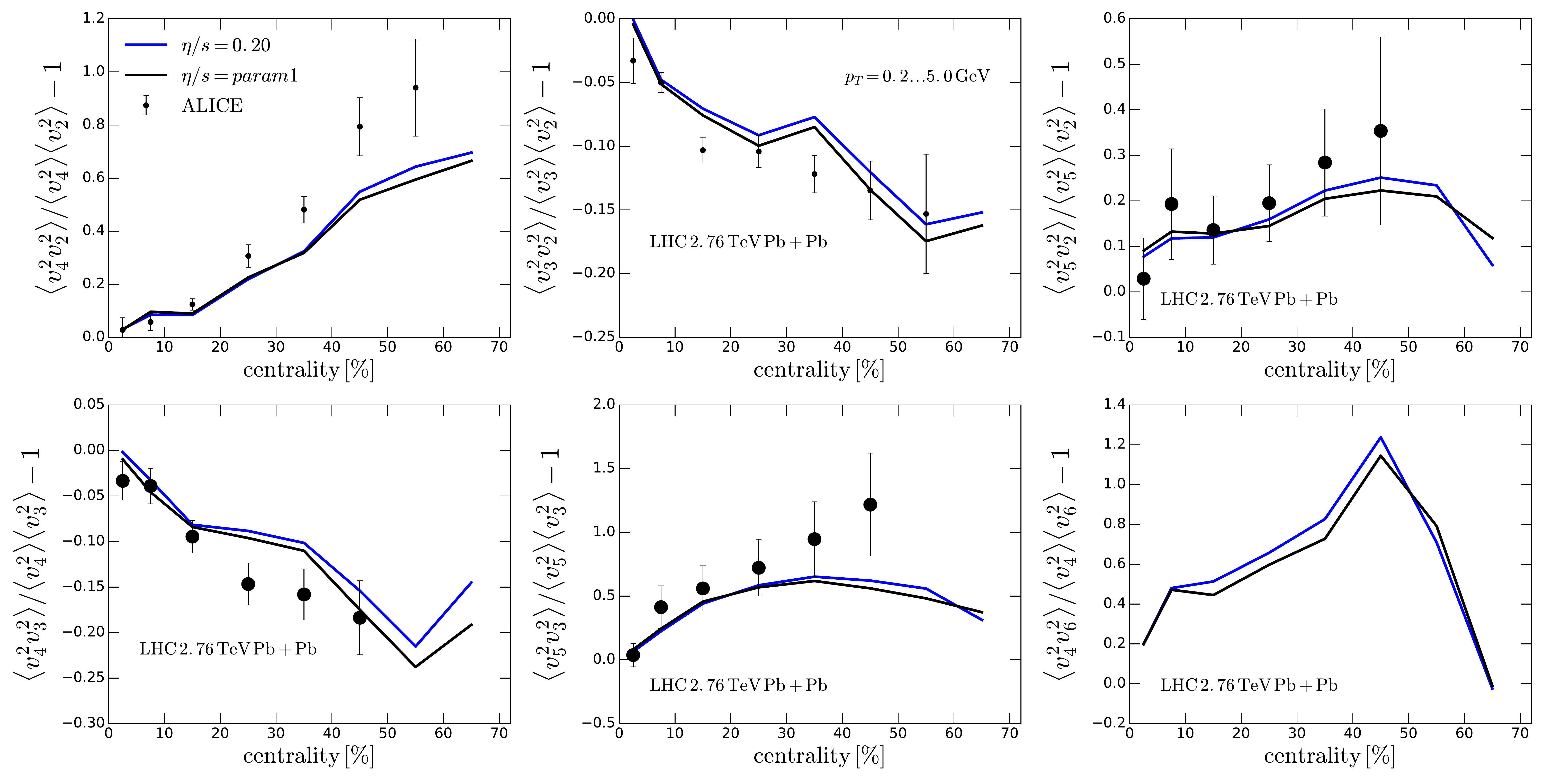}
\end{center}
\vspace{-0.5cm}
\caption{\small The normalized symmetric cumulants in 2.76 TeV Pb+Pb collisions. The data are from Ref.~\cite{Acharya:2017gsw}}
\vspace{-0.3cm}
\label{fig:nsc}
\end{figure}

In Fig.~\ref{fig:nsc} we show various correlator measures between $v_n$ and $v_m$, defined through so called normalized symmetric cumulants, $nsc(n, m) = \langle v_n^2 v_m^2 \rangle /  \langle v_n^2\rangle \langle v_m^2 \rangle -1$, where the angular brackets denote the event averaging, accounting the proper weights of the event multiplicity~\cite{Acharya:2017gsw}. The best agreement with the correlators is obtained with the same $\eta/s(T)$ parametrizations as the $v_n$ themselves, thus the consistency between the experimental data and the fluid dynamical behavior is excellent. 
%%%%%%%%%%%%%%%%%%%%% FIGURE %%%%%%%%%%%%%%%%%%%%%
%%%%%%%%%%%%%%%%%%%%% FIGURE %%%%%%%%%%%%%%%%%%%%%

\begin{figure}
\begin{center}
\includegraphics[width=0.55\textwidth]{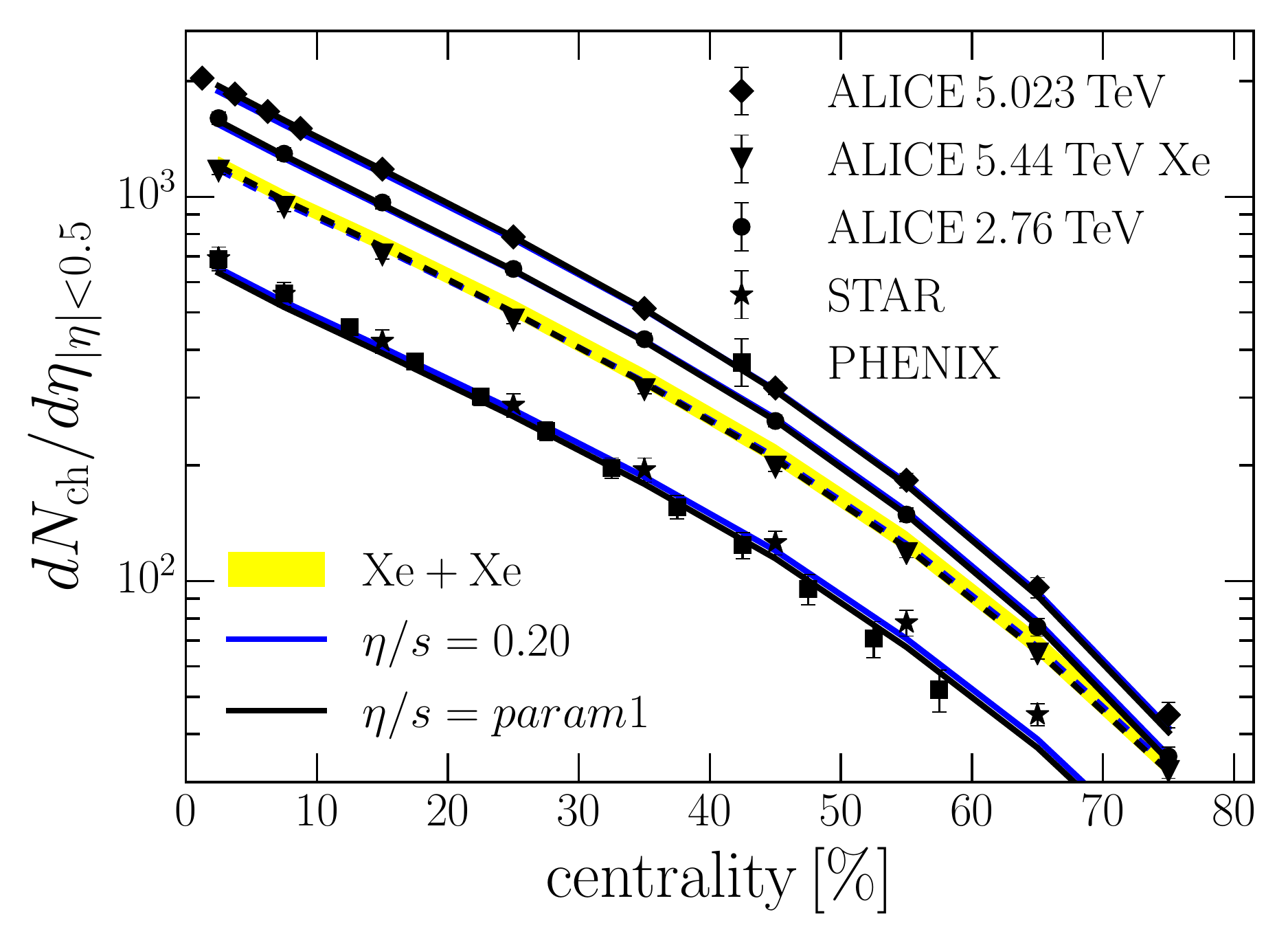}
\end{center}
\vspace{-0.5cm}
\caption{\small Centrality dependence of the charged hadron multiplicity. The experimental data are from ALICE \cite{Aamodt:2010cz, Adam:2015ptt, Acharya:2018hhy}, STAR \cite{Abelev:2008ab} and PHENIX \cite{Adler:2004zn}}
\vspace{-0.3cm}
\label{fig:multiplicity}
\end{figure}

\begin{figure}
\begin{center}
\includegraphics[width=0.9\textwidth]{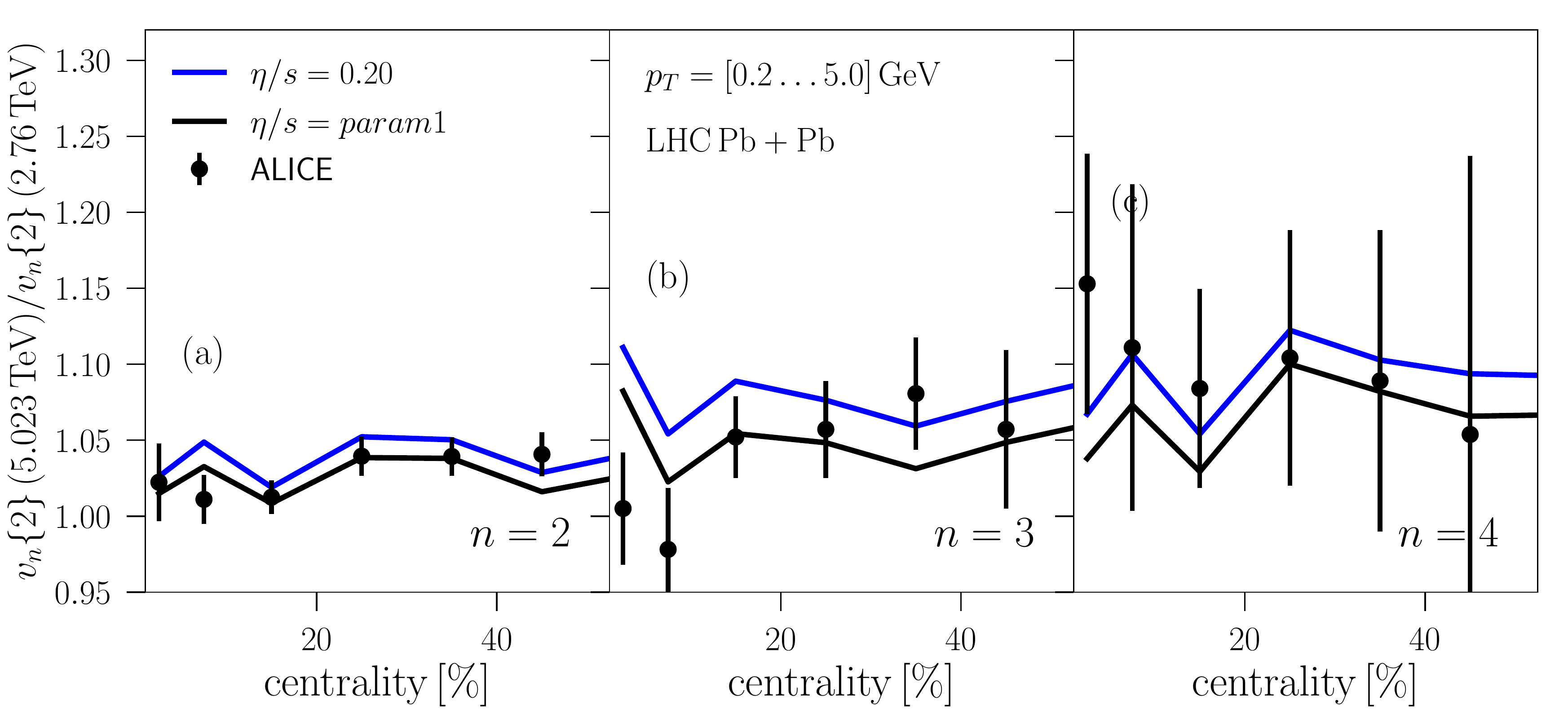}
\end{center}
\vspace{-0.5cm}
\caption{\small The ratio of $v_n$ between 2.76 TeV and 5.023 TeV Pb+Pb collisions. The data are from Ref.~\cite{Adam:2016izf}}
% \vspace{-0.3cm}
\label{fig:flowratio}
\end{figure}

Armed with the $\eta/s(T)$ parametrizations and the corresponding values of $K_{\rm sat}$ that were fixed in Ref.~\cite{Niemi:2015qia} our framework is closed, and we can predict the low-$p_T$ hadronic observables for 5.023 TeV Pb+Pb \cite{Niemi:2015voa} and 5.44 TeV Xe+Xe \cite{Eskola:2017bup} collisions that were measured recently. In Fig.~\ref{fig:multiplicity} we show the centrality dependence of the charged hadron multiplicity for all the systems we are considering. 
% The yellow band for Xe+Xe results indicates the uncertainty arising from the experimental error in $0-5$ \% 2.76 TeV Pb+Pb measurement of the multiplicity, to which $K_{\rm sat}$ is fixed. 
The predictions are in excellent agreement with the measured multiplicities.
The predicted $v_n$ in 5.023 TeV Pb+Pb collisions are shown in Fig.~\ref{fig:flowratio} as the ratios to $v_n$ in the 2.76 TeV Pb+Pb collisions. The data is from the ALICE Collaboration~\cite{Adam:2016izf}. The predicted slight increase of $v_n$ is consistent with the measurement. 
%%%%%%%%%%%%%%%%%%%%% FIGURE %%%%%%%%%%%%%%%%%%%%%
%%%%%%%%%%%%%%%%%%%%% FIGURE %%%%%%%%%%%%%%%%%%%%%

%%%%%%%%%%%%%%%%%%%%% FIGURE %%%%%%%%%%%%%%%%%%%%%
%%%%%%%%%%%%%%%%%%%%% FIGURE %%%%%%%%%%%%%%%%%%%%%

\begin{figure}
\begin{center}
\includegraphics[width=0.88\textwidth]{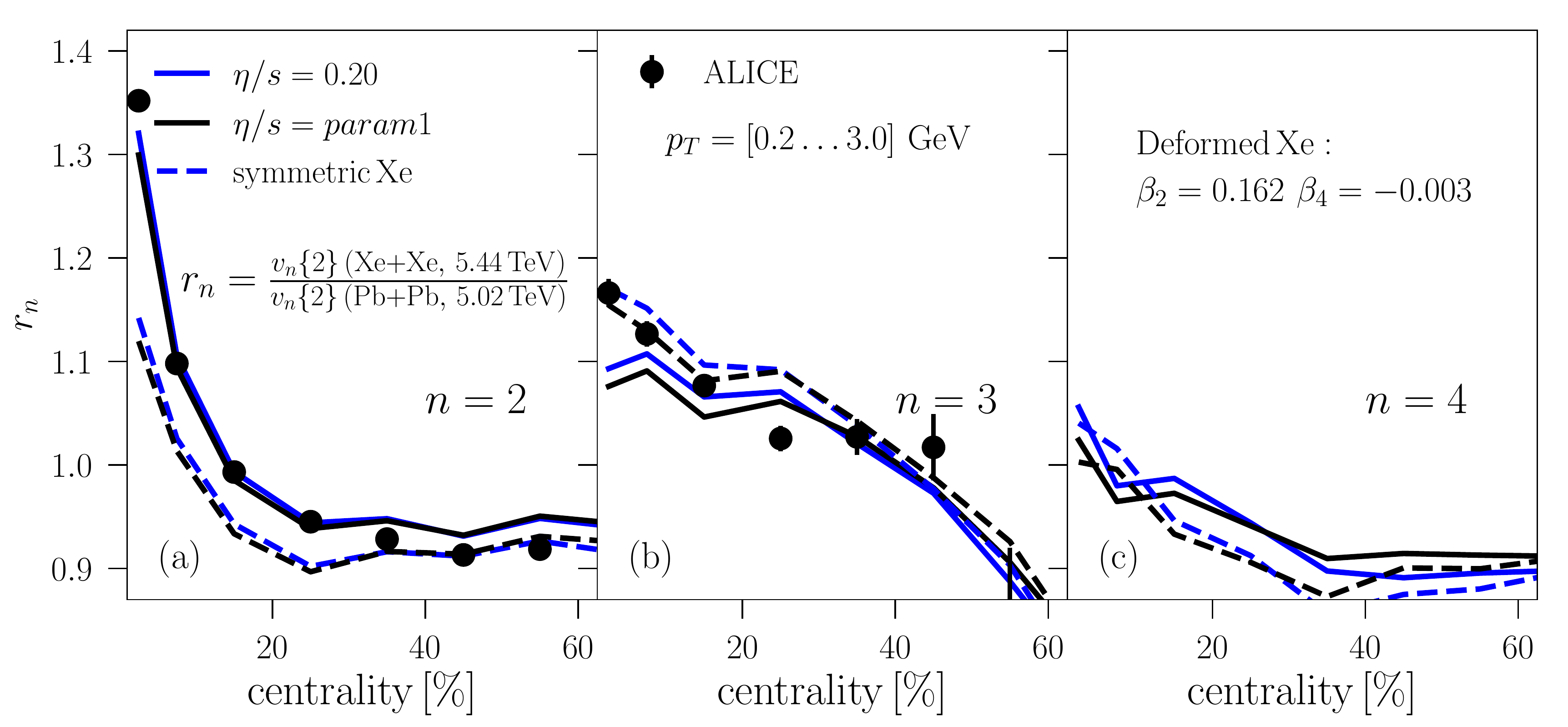}
\end{center}
\vspace{-0.5cm}
\caption{\small The ratio of $v_n$ between 5.023 TeV Pb+Pb and 5.44 TeV Xe+Xe collisions. The data are from Ref.~\cite{Acharya:2018ihu}}
\vspace{-0.5cm}
\label{fig:flowratio_Xe}
\end{figure}

Finally, in Fig.~\ref{fig:flowratio_Xe} we show the ratio of our predicted $v_n$ for the 5.44 Xe+Xe collisions to $v_n$ in 5.023 TeV Pb+Pb collisions compared to the ALICE data~\cite{Acharya:2018ihu}. The Xe nuclei are not spherically symmetric, and as a result the initial conditions are modified compared to the spherical case, especially in more central collisions, and moreover the modified initial densities then reflect through the fluid dynamical evolution to the final flow coefficients~\cite{Giacalone:2017dud}. In the figure we show both the cases: with and without accounting for the nuclear deformation. As can be seen from the figure, $v_n$ in central collisions are enhanced compared to the $v_n$ in 5.023 TeV Pb+Pb collisions regardless of whether we include the Xe deformation or not. However, with the deformation the enhancement of the elliptic flow, $v_2$, is much stronger, and it is clearly necessary to include the deformation in order to describe the experimental data. For larger $n$ the effect of deformation is much weaker.  
%%%%%%%%%%%%%%%%%%%%% FIGURE %%%%%%%%%%%%%%%%%%%%%
%%%%%%%%%%%%%%%%%%%%% FIGURE %%%%%%%%%%%%%%%%%%%%%

As a conclusion, we have demonstrated that our framework of EKRT initial conditions together with the dissipative fluid dynamical evolution describes a large class of low-$p_T$ observables in a wide range of collision energies. Especially, we have shown that once the framework is fixed at 200 GeV Au+Au and 2.76 TeV Pb+Pb collisions, we can predict the $\sqrt{s}$ and $A$ dependence of the soft observables and describe all these systems with the same $\eta/s(T)$, which is a necessary requirement to demonstrate a fluid dynamical behavior. In particular, the measurement of the Xe+Xe collisions is a very good test of the $A$ dependence of the initial particle production and fluid dynamical behaviour. The measured $v_n$ show exactly the behaviour we expect from the fluid dynamical response to the modified geometry due to the nuclear deformations.         

{\bf Acknowledgments} This work is supported by the Academy of Finland, Projects 297058 and 310130, and by the European Research Council, grant no. 725369. We acknowledge the CSC -- IT Center for Science in Espoo, Finland, for the allocation of the computational resources.



%% References


% \bibliographystyle{elsarticle-num}
% \bibliography{<your-bib-database>}


% \vspace{-0.2cm}

\end{document}